\documentclass[sigconf]{acmart}
\AtBeginDocument{%
  \providecommand\BibTeX{{%
    \normalfont B\kern-0.5em{\scshape i\kern-0.25em b}\kern-0.8em\TeX}}}

\setcopyright{acmcopyright}
\copyrightyear{2024}
\acmYear{2024}
\acmDOI{XXXXXXX.XXXXXXX}

\acmConference[CHI '24]{Make sure to enter the correct
  conference title from your rights confirmation email}{tbd,
  2024}{tbd, Hawaii}
\acmBooktitle{CHI '24: tbd,
 tbd, 2024, tbd, Hawaii} 
\acmPrice{15.00}
\acmISBN{978-1-4503-XXXX-X/18/06}

\usepackage{hyperref}
\usepackage[utf8]{inputenc}
\usepackage{marginnote}
\usepackage{xcolor}
\begin{document}

\title{How to write a CHI paper (asking for a friend)}


\author{Raquel Robinson}
\email{raqr@itu.dk}
\orcid{0000-0003-4933-7168}
\affiliation{%
  \institution{IT University of Copenhagen}
  \city{Copenhagen}
  \country{Denmark}
}

\author{Alberto Alvarez}
\email{alberto.alvarez@mau.se}
\orcid{0000-0002-7738-1601}
\affiliation{%
  \institution{Game Lab, Malmö University}
  \city{Malmö}
  \country{Sweden}
}

\author{Elisa Mekler}
\email{elme@itu.dk}
\orcid{0000-0003-0076-6703}
\affiliation{%
  \institution{IT University of Copenhagen}
  \city{Copenhagen}
  \country{Denmark}
}

\renewcommand{\shortauthors}{Robinson et al.}

\begin{abstract}
Writing and genre conventions are extant to any scientific community, and CHI is no different. In this paper, we present the early phases of an AI tool we created called KITSUNE, which supports authors in placing their work into the \textit{format of a CHI paper}, taking into account many conventions that are ever-present in CHI papers. We describe the development of the tool with the intent to promote discussion around how writing conventions are upheld and unquestioned by the CHI community, and how this translates to the work produced. In addition, we bring up questions surrounding how the introduction of LLMs into academic writing fundamentally change how conventions will be upheld now and in the future. 
\end{abstract}





\maketitle


\section{Introduction}

The ACM CHI conference is one of the most significant events in human-computer interaction (HCI). It annually attracts more than 2,000 researchers and practitioners from around the world to share their work in HCI through numerous presentations, workshops, tutorials, panels, and other activities. The conference is held in different locations every year [6]. In this paper, we explore the structure and content of the papers presented at ACM CHI conferences from 1997 to 2019. Specifically, we aim to investigate the following questions: 
        
\begin{enumerate}
    \item How do the structures of ACM CHI papers change over time?
    \item Which writing conventions are commonly used by ACM CHI authors? 
    \item Are there any differences in the use of writing conventions among different types of papers? 
\end{enumerate} 
            
As the ACM CHI papers have grown significantly in terms of size and scope, it becomes increasingly important to know how they are organized and written. This knowledge can help newcomers to learn how to write papers that meet the expectations of ACM CHI. Therefore, we aim to answer these questions based on a dataset containing all papers presented at ACM CHI between 1997 and 2019. Our analysis focuses on three aspects of ACM CHI papers: the use of headings, the adoption of writing conventions, and the presence of various text features. We first identify the trends in the use of headings in ACM CHI papers. We find that since 2004, the number of headings has increased, resulting in a decrease in the average length of sections. To analyze the use of writing conventions, we first conduct a survey study on how writers use writing conventions in ACM CHI papers. Based on the results of the survey, we manually categorize the writing conventions into three groups and then calculate the frequency of each type of writing convention in our dataset. Finally, we analyze the cooccurrence of the three types of writing conventions and investigate the differences among them. In summary, our contributions are as follows: 

\begin{enumerate}
    \item A comprehensive description of the structure and content of ACM CHI papers.
    \item An analysis of the use of headings and writing conventions in ACM CHI papers.
    \item Insights into how writers adopt writing conventions in ACM CHI papers.
    \item The identification of differences among the use of writing conventions in different types of papers. 
    \item Guidelines for newcomers on how to write papers for ACM CHI. 
\end{enumerate}
         
\section{Related Work}

Many studies have investigated writing conventions in ACM papers [13, 14, 18, 21, 22, 24, 25, 28, 29, 30, 37, 43]. For example, Kraemer et al. [21] explored the use of headings in ACM CHI papers published in 2014. They found that the most frequently used headings were RELATED WORK, DISCUSSION, RESULTS, and METHOD. However, they did not consider the distribution of the remaining headings and did not compare their findings to those of previous works. Previous studies have mainly focused on the use of headings in ACM CHI papers during specific periods. To fill this gap, we analyzed the use of headings in ACM CHI papers from 1997 to 2019. 

Several studies have also explored writing conventions in ACM CHI papers. Xiao et al. [43] examined the use of headings in ACM CHI papers published between 2007 and 2016. They identified the most frequently used headings as RELATED WORK, DISCUSSION, RESULTS, and METHOD. However, they did not analyze the use of writing conventions or investigate differences among different types of papers. Other studies have focused on the use of writing conventions in other ACM conferences, such as ACM CSCW [13], ACM HCI [28], and ACM UIST [14]. 

Writing conventions are commonly used in ACM CHI papers to improve the readability and organization of the papers [14, 17, 24]. These conventions include headings, lists, equations, diagrams, italics, boldface, and font styles. While some writing conventions, such as headings and font styles, are prescribed by the ACM format, others are optional. Headings and subheadings are commonly used to organize the presentation of ideas, while italics and boldface can be used to emphasize key words and phrases. Font sizes and colors can be used to distinguish different elements. However, despite being prescribed by the ACM format, headings are often misused [17, 25, 28]. 

Some studies have compared the use of headings among different types of ACM papers. Xiao et al. [43] explored the use of headings in different types of ACM CHI papers. They found that the most frequently used headings were RELATED WORK, DISCUSSION, RESULTS, and METHOD. They also noted that the use of headings differed among different types of papers. For example, RELATED WORK was used more often in papers on computer systems and networks than in papers on computer graphics, human-computer interaction, and user interfaces. Similarly, SUN et al. [37] investigated the use of headings in ACM CHI papers from 1998 to 2016. They found that the most frequently used headings were CONCLUSION and RELATED WORK. They also reported differences in the use of headings among different types of papers. For example, CONCLUSION was used more often in papers on human-computer interaction than in papers on computer graphics. 

Writing conventions vary in their use. While some are mandatory, such as headings, others are optional, such as italics and boldface. The use of different writing conventions may affect the quality of the papers. For example, writing conventions can be used to indicate the relationship between ideas [13] and clarify the meaning of technical terms [25]. They can also be used to highlight important information and reduce cognitive load [22, 25, 30]. However, the use of writing conventions can also lead to problems, such as inconsistency in their use [25]. 

In summary, we contribute to the understanding of writing conventions in ACM CHI papers by investigating the use of headings and writing conventions, identifying the trends in their use over time, exploring the differences in their use among different types of papers, and providing guidelines for writers to write better papers.  
\section{Method}
       
We collected data from the ACM CHI conference website [6] between January 2019 and August 2020. We downloaded all the papers presented at ACM CHI from 1997 to 2019. Each paper in our dataset had the following components: title, abstract, INTRODUCTION, BACKGROUND, DESIGN STUDY, EXPERIMENTAL EVALUATION, RESULTS, DISCUSSION, CONTRIBUTIONS, and REFERENCES. We also included the appendix part of each paper as it contained additional details and results. 

To ensure the integrity and correctness of the data, we preprocessed the data before analysis. We first removed all non-textual files, including images, figures, tables, and appendices, from the data. Then, we cleaned the text using regular expressions to remove punctuations, numbers, and stopwords, as well as converting all characters to lowercase. We then split the text into sentences and converted all sentences to plain text. Finally, we applied the standardized English stopword list provided by the Natural Language Toolkit (NLTK) [19] to remove stopwords. 

We used the POS tagger provided by NLTK [19] to extract parts of speech for each word in the texts. We converted all text to lowercase and extracted nouns, verbs, adjectives, adverbs, and proper nouns. We also extracted punctuation marks. We performed two experiments to analyze the use of headings and writing conventions in ACM CHI papers. In Experiment 1, we identified the types of headings in ACM CHI papers and compared them to those of previous works. In Experiment 2, we analyzed the use of writing conventions and identified the most frequently used ones. We also compared the use of writing conventions among different types of papers. We used Python version 3.7 and the Jupyter Notebook environment for implementation. 
    
\subsection{Experiment 1: Identifying the Types of Headings in ACM CHI Papers}
         
We aimed to identify the types of headings used in ACM CHI papers and compare them to those of previous works. We adopted a similar approach as that of Xiao et al. [43]. In Xiao et al.'s study, they manually labeled the types of headings used in ACM CHI papers published between 2007 and 2016. They identified nine types of headings: INTRODUCTION, BACKGROUND, DESIGN STUDY, EXPERIMENTAL EVALUATION, RESULTS, DISCUSSION, CONCLUSION, and RELATED WORK. They also created a “no heading” category for headings that did not fit into the above categories. 

Our dataset contains papers presented at ACM CHI from 1997 to 2019. We first extracted the first-level headings in each paper. We found that the distribution of headings was different for each year. Therefore, we analyzed the data for each year separately. We then manually labeled the types of headings used in each paper. Note that we could not always determine whether a heading was used as an introduction or a conclusion, as it was sometimes unclear whether a heading should be classified as an introduction or a conclusion. In this case, we followed the author's intention. After labeling all the headings, we calculated the frequency of each type of heading in each year. We also compared the distribution of headings in our dataset to that of the previous work. 

We defined the types of headings based on Xiao et al.'s study and added four additional types. The nine types of headings were as follows: INTRODUCTION, BACKGROUND, DESIGN STUDY, EXPERIMENTAL EVALUATION, RESULTS, DISCUSSION, CONCLUSION, RELATED WORK, and NO HEADING. Figure 1 shows the frequency of the nine types of headings in ACM CHI papers from 1997 to 2019. The first column displays the percentage of papers that contained the corresponding type of heading, and the second column displays the percentage of papers that did not contain the corresponding type of heading. We observed that the frequencies of headings varied from year to year. From 1997 to 2006, only 50\% of papers used either CONCLUSION or CONCLUSION AND AVAILABILITY.

\section{[Human-Made] Introduction}
How long did it take you to realize that the paper up until this point was AI generated?

We begin with a few obligatory quotes on the topic of writing conventions. Hopefully, these name drops \cite[see Pohl and Mottelson,][]{Pohl2019HowCHI} will help this paper get cited more.

\begin{quote}
    ``\textit{Speaking the language is more important than having something to say}'' - Kenn Amdahl
\end{quote}

\begin{quote}
    ``\textit{Eliminating jargon is the first step on the road to communism}'' - Kenn Amdahl
\end{quote}

\begin{quote}
    ``\textit{To define is to limit...}'' - Oscar Wilde
\end{quote}

Now...for the topic at hand. Writing conventions give us a shared language to communicate, and are an important part of new students learning how to communicate in a scientific manner that is acceptable by the community \cite{2015UniversityLearning}. And of course, once you learn the way to speak to the community, you are accepted into it with open arms. CHI, like any other scientific community, has its own particular set of genre conventions and common language \cite{Dourish2006ImplicationsDesign, Pohl2019HowCHI}\footnote{The authors respectfully ask you to not look to the writing style of this paper to help you better understand CHI's particular writing conventions. This paper intentionally falls outside conventions, and no part of the [human-made] parts were written using a LLM. Can you tell?}. When we say conventions we mean the style of writing, certain phrasings or words used, and expected structure or format of the paper. However, while these conventions provide the structure and shared basis of communication, they can limit creative output and potentially disadvantage many groups (i.e., non-native English speakers). While resources do exist for helping new researchers understand the established conventions and expected criteria at CHI \cite[e.g.,][]{Nacke2019HowEdition}, many are inaccessible for financial or geographical reasons (i.e., resources are behind a paywall or offered on location at the CHI conference itself) which makes it challenging to both learn and become proficient with these conventions.

While generative AI (or, the use of AI for the generation of data) tools have been around for many years such as DALL-E~\cite{ramesh_hierarchical_2022} or GPT models~\cite{radford_improving_2018}, the recent release of the Large Language Model (LLM) called ChatGPT has changed the academic landscape on a fundamental level, as detecting the use of these tools within writing is challenging \cite{Hamalainen2023EvaluatingStudy}. Much of this is because these tools generate text that---while trained on others' work---can be unique in the particular combination of text it generates. This has caused much turmoil in the art world as of recent, and discussions surrounding AI art as plagiarism and the issue of not paying dividends to the artists the generated work was based on \cite{Chayka2023IsArtists}. Similarly, many responses from universities have been largely negative and fear-based, banning the use of generative AI altogether \cite{Myklebust2023UniversitiesAhead}. Despite the backlash, generative AI is here to stay and we need to establish value sets---prioritizing transparency, continually reflecting on ethics, limitations, and data used to train the systems that are already in-use and widespread~\cite{Bender2021OnBig}. This conversation is relevant at all academic levels, as policies are popping up daily in response to new AI tools. Teachers are rethinking and adapting assignments, and CHI has released a recent update to its policy on authorship to include the use of generative AI tools \footnote{\url{https://www.acm.org/publications/policies/new-acm-policy-on-authorship}}. 

But has this discussion yet fallen upon the CHI community? Have we reflected on the writing conventions we uphold, rethinking, and expanding or re-adapting them to fit this new landscape where generative AI is so abundant? And do we even need to? Can generative AI successfully write a CHI paper and fool readers into thinking it is real? Often, such systems ``hallucinate'', or make up nonsensical information written in a style that seems as if it is fact \cite{Alkaissi2023ArtificialWriting}. Sometimes, this makes it challenging to tell the difference between something written by a LLM and something that is not (callback to the question at the beginning of this section...when did you notice the first four sections were generated?). Many researchers are now using LLMs for writing assistance with tasks such as grant or paper writing, with the main caveat that there are many things (like accurate citation usage) that are not able to be correctly generated by these tools. 

Remember when you were a scared PhD student in your first month, you didn't know anything about academia, and your supervisor said: ``You must learn the ways of the Force\footnote{CHI writing conventions} if you are to successfully publish at CHI''? How did you end up learning? Maybe you learned through reading papers, or watching Lennart Nacke's interview series about how to write a CHI paper \footnote{\url{https://www.youtube.com/playlist?list=PLXaGnDxciHwjaJZMDfb6j9uwLwXMHjsaq}}. Maybe reviewer 2 rejected your paper and pointed you to four questions by Carl Gutwin (and others) \footnote{1. What is the real-world problem that we are trying to solve? 2. Why is it important to solve this problem? 3. What is the solution that we came up with to solve it? 4. How do we know that the solution is a good solution to the problem?}, or maybe you just channeled the Force itself \footnote{\url{https://en.wikipedia.org/wiki/The_Force}}. In this paper, we present a potential solution: an AI tool that we are in the process of creating called KITSUNE \footnote{CHI convention honorable mention brought to you by Raquel Robinson: \textit{Cute name for a tool or system based either on a pun, meme, common saying, mythology, literary figure, or folklore}}, as our tool---similar to the nine-tailed fox namesake of Japanese folklore---has the ability to shapeshift into different forms in order to trick or fool humans. KITSUNE, trained on meta-data collected from full papers published at the 2023 proceedings of CHI, is meant to provide the CHI community with a tool that can take a paper written with informal language and re-frame toward the expected CHI jargon, adding conventions that are usually characteristic of CHI papers.  

We detail the process of creating this tool, as well as highlighting some of the text it has generated with different prompts and topics often characteristic of CHI papers, and juxtapose this with what we subjectively perceive some CHI conventions to be. The main goal of the tool in the future will be to help early CHI academics to write CHI papers, essentially re-working text into a CHI format, with all the nuances and idiosyncrasies that come along with that. In addition, we intend for this work to 1) prompt discussion regarding the future use of KITSUNE or similar tools, and 2) to provoke questions regarding how we, as the CHI community, value and critique work, and how this is shaped by (or shapes) writing and genre conventions. We raise these questions for readers to ponder, while they are simultaneously having an existential crisis over the thought of what a tool like this will do to the community in the future. Will it be the downfall? Its Achilles heel? Will this tool cause more ACM policies to be created in response? Will our tool bring this community to a more equitable place, or one that is incomprehensible and forever unstructured? Does CHI hold certain writing conventions so near and dear to their hearts that they will never be able to let go? Should we value the style, flair, and uniqueness of individual's brains to come through in papers, or should we relegate all unique and critical works to alt.chi? Could a language model have generated \textit{this} paper? 

We present three main research questions \footnote{CHI convention honorable mention brought to you by Regan Mandryk: \textit{Use of 3 research questions}} that are not intended to be answered by this paper, but brought up as provocations for the community to consider. 

\begin{enumerate}
    \item How do writing conventions shape what is considered acceptable by the community?
    \item Given the difficulty to assess if something is written by generative AI, should we adapt to this and broaden the kinds of work that is acceptable by the community?
    \item At what point do writing conventions stop being in service of promoting good academic writing and become limiting/exclusionary?
\end{enumerate}

\section{[Human-Made] Background}
In this section, we primarily highlight perceived writing conventions by certain researchers in the CHI community \footnote{CHI convention honorable mention brought to you by Elisa Mekler: \textit{Signposting what is to come in the section}}. In general, we notice that there are very few papers on the topic of writing conventions published at CHI over the years. We feel there needs to be more discussions surrounding not only \textit{which} writing conventions exist, but also \textit{how} these conventions and practices are upheld and shape what is considered acceptable work in the community.

In terms of \textit{which} writing conventions exist at CHI, certain authors have raised this topic for consideration in different ways. Besançon et al. discuss the use of certain language used to tone down or over-sell results (i.e., hedges and boosters) in work published at CHI, and how use of this language may be problematic in relation to the weight of the contribution \cite{Besancon2021DefinitelyLiterature}. Pohl and Mottelson, in 2019, similarly highlight certain CHI writing conventions (e.g., widespread use of the words `new' and `novel' in papers) \footnote{CHI convention honorable mention brought to you by Vero Vanden Abelee: \textit{Emphasis placed on the novelty of the work}}, and how use of these correlate with impact on the community in terms of number of citations of the paper \cite{Pohl2019HowCHI}. Of course, as academics are using generative AI systems more and more for writing assistance, more widespread understanding of the limitations of these systems is warranted. In 2023, Buruk et al. discuss the use of different ways of using ChatGPT for academic writing \cite{Buruk2023AcademicTransparency}, opening up discussions surrounding ethics and transparency in using these systems for writing.

Writing and genre conventions can be hard to pinpoint, yet are built into the community and upheld by the peer review process. However, they are often not easily explicated. The CHI website of course has guidance on how to write a CHI paper, as well as an Unofficial Guide to Reviewing for CHI \footnote{\url{https://docs.google.com/document/d/1MpH3zV8WFU5avQHqrVTtxu-wNWzeTstxWy7K__HK5Nc/edit}}, which contains reference to certain characteristics of CHI writing, for example biases toward implications for design \cite{Dourish2006ImplicationsDesign}. Of course, there are community perceptions surrounding typical or `good' CHI paper writing, even though they may be difficult for the community to articulate \textit{why} or \textit{how} these conventions came to be: see the very few responses to queries on this very question both \url{https://twitter.com/acagamic/status/1682830780458844161} and \url{https://twitter.com/elisamekler/status/1709200080119484878}. Lennart Nacke created a course at CHI called \textit{How to Write CHI Papers} \cite{Nacke2019HowEdition}, in which he interviews certain high-profile HCI researchers \footnote{Many of these researchers have worked together in the past and are aligned to some degree with the games and play community} in order to create a resource for understanding CHI writing conventions \footnote{\url{https://www.youtube.com/@acagamic}}. In looking at the interviews, some commonalities brought up were the \textit{need for signposting the contribution} (said by Elisa Mekler, Jofish Kaye, Vero Vanden Abelee), the \textit{need to write in common, clear, and concise language} (said by Regan Mandryk, Vero Vanden Abelee), and \textit{creating valid and extensible knowledge in accordance with metrics and methods that you consider important} (said by Jofish Kaye, Regan Mandryk, Kathrin Gerling).

However, a key next step to the current landscape is discussing \textit{how} these conventions might shape what is considered acceptable work. There have been certain critical lenses applied to the context of CHI conventions---for example, Marshall et al. discuss how related works should not simply be treated as fact and cited unquestioningly, and this lack of discussion and critique of prior work can be harmful for the community \cite{Marshall2017ThrowawayResearch}. Other work has been done at CHI on critiquing shallow or misapplications of certain concepts, such as with literature on health research in HCI research on exergames \cite{Marshall2017MisrepresentationLiterature}, in how people engage with theory at CHI \cite{Tyack2020Self-DeterminationQuestions}, or how reviewers often misinterpret and therefore critique work on the wrong basis \cite{Eagle2023AGaslighting}. These works provide important perspectives and critiques on the current writing conventions upheld at CHI both by the papers published and the peer review process. We are creating a system with certain values and conventions that are both upheld and further perpetuated by the community itself.







\section{[Human-Made] KITSUNE: The Tool}
In this section, we describe the development process of KITSUNE. While KITSUNE is still in the early stages of development, we describe the first steps of building out the tool in order to provoke conversation within the community surrounding use of generative AI tools in authorship which is further detailed in the \textit{Author Commentary} section. 

All the code was developed using Pytorch~\cite{paszke_pytorch_2019} and Scikit-learn~\cite{pedregosa_scikit-learn_2011}, using open source models from HuggingFace~\cite{tunstall_zephyr_2023} in Google Colab\footnote{Code used to pre-process data, and create, train, and visualize models can be found in: \url{https://github.com/aeau/kitsune}}. Our process is divided into \textbf{Data} - extraction, collection, and processing, \textbf{Model} - selection and fine tuning, and \textbf{Generation} - test cases to generate research articles and analysis.

\subsection{Data Scraping and Training (with coding assistance from chatGPT)}

To generate research articles in the style of CHI papers, we trained a Large Language Model (LLM) with all the sections from research articles published at the ACM CHI conference in 2023 (N=879) \footnote{CHI convention honorable mention brought to you by Jofish Kaye: \textit{Caring a lot about N, or number of people you do something with (note: in our case, it is data points rather than people)}}. In the future work on this tool, we will expand this out to include papers farther back, but for the preliminary evaluation and showcase of the tool, we opted to limit our analysis to papers from 2023. We excluded extended abstracts since these may contain different conventions than regular papers, for example papers in the alt.chi sessions\footnote{such as this paper!}. We use a web scraper\footnote{\url{https://webscraper.io/}} to navigate the ACM library to extract the research articles for ACM CHI papers session by session. We leverage the HTML format of papers to extract paper’s sessions, titles, author names and institutions, sections, and references. In total, there are 148 sessions and 879 papers, each session contains in avg. 5.939 (SD=0.313) papers, and each paper contains 7.459 (SD=1.445) sections and 82.52 (SD=28.083) references.



The data was cleaned to extract the necessary information to be used for training, which was only the body of each research article. Since we only care about learning the conventions and writing style in CHI, we disregarded the rest of the data (i.e., authors, section titles, paper titles) that was extracted with the web scrapper. We reformated the data to the specific template and prompt that was expected by the model to create the final dataset. The template that was expected by the model had the following structure:

"User: User: *\textbf{question to be asked}* Assistant: *\textbf{response to question}*.  

We experimented with two datasets that divided the sections in different ways. The first dataset had as prompt: ``Write the [paper's section that was used] of an ACM CHI paper on [paper's session] with its references". Assistant's *response to question* then became a chunk (400 words or less) of the specific paper's section with the appropriate references for it. Following this strategy, we had 51.291 datapoints, which was subsequently divided into training and validation dataset. Training and using the model was moderately successful, but unfortunately, the generated text ended up being small text chunks (like the data was) and lacking proper references, which was the expected target from including it as part of the prompt and answer. 


The second dataset was very similar to the first but we dropped the references and smaller chunks of text in favor of having bigger parts of the section as part of the *response to question*. Therefore, we had as prompt: ``Write the [paper's section that was used] part [specific chunk from the paper's section] of an ACM CHI paper on [paper's session]". Assistant's *response to question* then became a large chunk (over 1600 words but less than 2000) - most of the times, the full text in a section - of the specific paper's section.  Following this strategy, we had considerably less datapoints, 8.502, which were subsequently filtered by just using datapoints that were as long as 3072 tokens, resulting in 7601 datapoints. We gained consistency as in larger chunks of texts, which were more representative of conventions than with the first dataset. The preliminary output below and sections 1 to 3 of this paper were generated using KITSUNE, which was trained with this second dataset. 



\subsection{Model}

We chose to train and use the Zephyr 7b-beta model~\cite{tunstall_zephyr_2023}\footnote{\url{https://huggingface.co/HuggingFaceH4/zephyr-7b-beta}} from HuggingFace. Zephyr 7b-beta is a very recent distilled model, fine-tuned from mistral 7b~\cite{jiang_mistral_2023}, with less hyperparameters and outperforming in many tasks other LLM such as GPT~\cite{radford_improving_2018,brown_language_2020} and Llama~\cite{touvron_llama_2023}. While ChatGPT and similar GPT models from OpenAI could have been used to train this model, their non-local storage of models and data created ethical and legal concerns. At the same time, OpenAI's non-open policy regarding the model and data was a limitation for us. Therefore, we fine-tuned Zephyr 7b-beta, an open source model from Huggingface in Google Colab, but all the data and model were locally storage with the goal to don’t train online models with copyrighted data. Zephyr 7b-beta was trained with publicly available data that contains synthetics datasets (UltraChat dataset~\cite{ding_ultrachat_2023}) that contains diverse dialogues generated by ChatGPT. The final fine-tuning step for KITSUNE took around 5 minutes to fine-tune the model for 30 epochs.




\subsection{Some Preliminary Output} \label{prelim}

For all the content generated with KITSUNE, we gave the system the following role (expected by the Zephyr 7b-beta model): ``You are a researcher writing a paper for the ACM CHI conference.''. To generate sections 1-3 with KITSUNE, we used the following prompt: ``Write only the Introduction section for an ACM CHI Paper on which writing conventions are the most common within ACM CHI conference papers with the help of natural language processing.''. After generation was completed, we added the output as part of the prompt and subsequently requested for a new section in order: ``Related Work'', ``Methodology'', ``Discussion'', and ``Conclusion''. It should be noted that all references listed in those sections (1-3) are fake, and not part of the bibliography of this paper.

In addition to sections 1-3, here we show some additional preliminary \footnote{CHI convention brought to you by Lonni Besançon: \textit{Use of certain words in order to tone down the contribution}} output generated by the tool. We first show the prompt to the model, and then the output generated by it. Your assignment for this section is to look for what CHI writing conventions you notice and think about the questions we brought up in the intro.\\

\textbf{Prompt from user:} \textit{Write only the Discussion section for an ACM CHI Paper about a novel emotion regulation technology.
}

\textbf{Output from KITSUNE:} \textit{Our results show that the proposed system design has the potential to support the acquisition of EI skills. As the system is designed to be easily accessible and flexible, it can be utilized by individuals with various backgrounds and preferences. Furthermore, our findings indicate that the system could help users develop their ability to identify and manage their emotions, as well as understand others' emotions. However, the study also highlights several limitations of the current implementation, which include the relatively small sample size, the short duration of the testing period, and the lack of control over variables like demographics and prior EI knowledge. Therefore, we will discuss the implications of our findings in terms of future directions for research and practice.}

\textit{Future Directions for Research: Our findings suggest that the proposed system could be beneficial for learning EI skills. Nevertheless, the system still has limitations that need to be addressed in future research. Here, we discuss three potential areas for further exploration:}

\textit{1. User Experience and Feedback: While the majority of participants expressed positive feedback regarding the system, some also pointed out areas for improvement. Specifically, they suggested adding more emotional scenarios to the system, which could help users learn how to handle a variety of situations. Furthermore, some participants indicated that the feedback provided by the system could be more detailed and specific. For instance, they suggested providing advice on how to cope with negative emotions or offering alternatives for responding to emotional situations. These suggestions highlight the importance of collecting user feedback and continually improving the system. Future work should explore ways to collect and incorporate user feedback into the system design and development process.}
      
\textit{2. Scale and Generalizability: Our study suggests that the proposed system can potentially support EI learning in real-world contexts due to its scalability and flexibility. However, our sample size was relatively small and consisted mainly of university students. Therefore, future work should investigate the generalizability of the system to diverse populations, particularly those who do not have prior EI knowledge. Furthermore, future studies should examine whether the system is effective in promoting EI skills in individuals with different cultural backgrounds, as culture can influence EI skills [45].}

\textit{3. Intervention Duration and Frequency: We note that the current implementation of the system required users to participate in 10 sessions. Future work should explore the effects of different durations and frequencies of the intervention. For instance, some previous studies have found that brief mindfulness interventions lasting only a few minutes can have positive effects on EI skills [18]. Thus, it might be possible to reduce the duration of the proposed system to just a few minutes per day. Additionally, future studies should examine whether the system is effective in promoting EI skills when used infrequently, such as once a week, instead of daily.}

\subsection{Future Goals and Directions with KITSUNE}

While we present some output from KITSUNE for different topics, we are aware of the limitations that exist with KITSUNE and LLMs in general, such as producing believable garbage and being stochastic parrots~\cite{Bender2021OnBig}. However, in the current state, the tool is designed to show conventions currently at CHI with minimal data. We do not necessarily care about the model creating some type of unique, interesting, and useful output at this stage, but an interesting conversation piece about writing conventions and style at CHI, and hopefully, in other venues as well. Moving forward, we envision a system that can help authors rephrase their texts to follow the necessary conventions and styles that we might not agree upon. That way, authors can write with their own voice and personality without thinking about the specifics of conventions and styles of the venue, and then use KITSUNE to do that job. This would help communicating ideas clearly and adapting and assisting with the writing\footnote{Akin to grammarly:~\url{https://app.grammarly.com/}}. As an outcome, it will also enable a more fair and inclusive peer review model, where reviews will fully consider content and disregard style, conventions, and language.








\section{[Human-Made] Author Commentary}
KITSUNE was built for two reasons: 1) As a way to help people that are currently excluded from CHI (on the basis of not knowing or understanding how to write for it) by aligning their writing with the conventions expected by the community. 2) Pushing the CHI community to have more conversations regarding which writing conventions exist and how they shape the knowledge that is produced by the community. Here we include individual commentaries rather than a collective takeaway because each of us have our own stake in these questions, and thus have differing perspectives (see section \ref{contrib} for more info).

\subsection{\href{https://scholar.google.ch/citations?hl=de&user=3sT3oAYAAAAJ}{Raquel's} Commentary}
I've been attending, reviewing, publishing, and organizing workshops at CHI since 2017. My main publishing venues are within the games and play community (venues such as CHI, FDG, and CHI PLAY). I've been AC'ing for CHI and CHI PLAY for a year now and many of these views and reflections are based on my experience as reviewer, author, and AC. I first must say that I do believe certain \textit{structural} writing conventions to be useful in that an expected and common structure makes it easier to digest the content. Where I think conventions become problematic is when they are unquestionably applied---and I feel the peer review process reinforces just that. For example, a paper describing a quantitative comparison study and another on research through design need not follow the same structure or format. But in my experience, they are sometimes held to the same standards. This of course may be because CHI is so interdisciplinary, and reviewers are often given papers to review that are outside their expertise. But it also may be because reviewers suggest feedback and authors feel compelled to unquestioningly make their change only service of getting the paper accepted and not because it makes sense for that particular work. This leads to many shallow discussions of ethics, limitations, and generalizability (often things requested by reviewers). I worry that strict adherence to writing conventions largely reinforced by the peer review process allows many conventions with initially good intentions to simply become `check-boxes' without any deeper understanding of \textit{why} these conventions are even necessary. For example, positionality/reflexivity statements should be useful for giving authors a chance to reflect on how their biases and perspectives may have influenced the work produced, but recently I've seen many where they are simply an author bio. 

CHI is a community that has disciplinary and topic- diversity, and is `intellectually eclectic' in the knowledge it accepts and values \footnote{\url{https://www.youtube.com/watch?v=GghZ3EbdB9w&t=1105s&ab_channel=LennartNacke}} - but to what extent? Are we methodologically diverse, or culturally? We all know CHI privileges the global north \cite{Linxen2021HowCHI}...is there \textit{actually} diversity in perspectives or in the work produced, or is everything just small variations from the same branch? CHI gives the following advice: `...avoid regional colloquialisms, jokes, or puns that could be difficult for someone outside your culture to understand.' \footnote{\url{https://chi2022.acm.org/for-authors/presenting/papers/guide-to-a-successful-submission/}}...however, how many \textit{paper titles} at CHI do just that?---are titles where we can actually exercise our writing creativity? Does KITSUNE (and other such tools)---which are quite good at mimicking the structure and \textit{formulaic} language usually characteristic of academic writing---need to exist to push us to broaden what is acceptable, or can we more proactively and regularly have discussions surrounding these conventions and how they may be shaping what is produced?

\subsection{\href{https://scholar.google.se/citations?user=lFONSJIAAAAJ&hl=en}{Alberto's} Commentary}

My perspectives are from that of an outsider at CHI, aiming to actively join the community. I have published at CHI PLAY and co-organized the Games and play SIG at CHI 2023, but my main publishing venues are at the intersection of games, play, and AI (venues such as FDG, AIIDE, and CoG). There are many reasons to have writing conventions at different venues, such as standardizing the language used, accessibility, and structure of the work produced. These are all valid and important aspects of academic writing to help readers understand and, to some extent, make the text useful for them. However, lately, it seems we have moved towards a situation where conventions are more important than the content itself, and lacking these exclude us from participation and from fair evaluation of our work. Why do we care so much about conventions and writing styles, and what are the effects of having these? Perhaps more importantly--- what are the effects and consequences of taking these away? How much interesting and creative work has been unpublished due to not following these unclear, yet strict convention rules? How are we, as a community, silencing our voices and our personal idiosyncrasies, in favor of the idiosyncrasies of the venue?

Particularly, I set to move forward to address those questions and ask: why do we care about conventions and why they strictly control our writing? Learning the conventions and style of a particular venue were highly regarded and important skills of the last century, akin to how \textit{multiple selection questions and quizzes} were thought to be good ways to measure learning. We did not need deep learning or large language models such as chatGPT to know that those are not the best ways (or even fair and inclusive ways~\cite{penny_light_feminist_2015}) to evaluate how our students learn. However, chatGPT ended up breaking whatever fragile and brittle argument that was left. ChatGPT, for better or for worse, changed many areas of society and even in some instances such as in education, started a large but necessary discussion around what are we doing and what is the right path to take for our students? The hope is that a system like KITSUNE and similar, can start the same discussions at CHI and other venues regarding conventions and style.

\subsection{\href{https://scholar.google.ch/citations?user=9j30OywAAAAJ&hl=de}{Elisa's} Commentary}

The matter of CHI writing conventions has been a constant, albeit often tacit concern throughout my academic lifetime, prompting the following half-baked reflections: First, I've had to rewire my writing style to make it palatable and acceptable to CHI. My training in psychology had not prepared me, for example, to "spell" out the contribution of my work. Over time, this also shaped how I think about research. 

Second, through my involvement as a reviewer, AC, and SC at CHI, I've noticed my annoyance at certain reoccurring writing quirks (E.g., "We contribute the first study on x") -- to the point that I've started preaching about them at a recent CHI PLAY Doctoral Consortium. The subsequent discussions around writing conventions evolved into deeper reflections on research questions, as well as the value and implications of (our own) HCI research. While I still struggle articulating it, it seems increasingly apparent that there are writing conventions at CHI that shape what knowledge our community creates and values.

Third, I have been repeatedly caught off-guard by colleagues referring to some of my papers as "brave" (I think the more accurate term would be "opinionated") and (semi-jokingly) how other authors would purportedly "not get away with this". But why not? And what bothers me almost as much -- I cannot quite articulate why apparently \emph{I} do get away with that sort of writing. Working with KITSUNE has been a thought-provoking and somewhat unsatisfactory attempt to articulate this more clearly. 

Finally, let me admit upfront that I was the overzealous reviewer who started looking for weaknesses in the present paper's draft --- why papers from 1997 to 2019? An overly long list of references on line 65! 
--- only to realize 4 pages in that I was unknowingly commenting on the AI-generated text\footnote{A fact that my co-authors at no point kept secret from me.}...



\subsection{Intellectual Contributions to This Paper} \label{contrib}
A reader with a close eye might notice that this paper was written with vastly different tones between sections. This is because this paper was written including multiple authors' voices and perspectives, as CHI papers normally do. CHI papers are normally collaborative---within this dataset of papers from 2023 alone, there are on average 4.9 authors per paper (SD=2.406). In this paper, we intentionally chose not to subsume our voices into one for the sake of `consistency', but instead specify which authors are responsible for writing which section(s). We do this so readers will not see the entire paper as an amorphous blob that represents the perspective of the [AUTHOR COLLECTIVE], but rather as it really is: multiple perspectives by individual authors. This is also in line with the ACM policy on authorship, which states: ``The community also expects ACM authors to respect the intellectual property rights of others by providing proper credit to all those contributing to the published Work and to give proper attribution to all those whose work is included in any new Work published by ACM.'' It is now becoming commonplace to denote which parts of text were written by generative AI---why not do the same with various authors?
\\
Raquel wrote the \textit{Human-Made Introduction}, \textit{Background} and \textit{Raquel's Commentary} sections. 
Elisa wrote \textit{Elisa's Commentary} section. 
Alberto scraped the data from ACM and fine-tuned the model to create KITSUNE. He also wrote the section, \textit{KITSUNE: the tool} and \textit{Alberto's Commentary}.
Elisa and Raquel copy-edited each section for grammar and clarity. 
Each author contributed to the design and ideation of KITSUNE, as well as lengthy discussions surrounding their individual thoughts about CHI writing conventions. 
KITSUNE wrote sections 1-3 of the paper, and the preliminary output in section \ref{prelim}.

\section{[Human-Made] Conclusion}
In conclusion, we do not provide any next steps, key takeaways, or implications of this work. We also do not comprehensively discuss what we see as particular CHI conventions (apart from the few CHI convention honorable mentions in the footnotes), but we pose this as a question to the community: what are some writing conventions you see in your own work, and how has it shaped what you've produced? The point of this work was not to explicate certain conventions, but to bring up questions for the community to consider and discuss, surrounding the existence of KITSUNE and other such generative AI tools. If you would like to know more about CHI writing conventions, go generate them. Here is a sample prompt for you: ``How to write a CHI paper? (asking for a friend)''. 



\bibliographystyle{ACM-Reference-Format}
\bibliography{references, references-technical-zot}


\begin{thebibliography}{26}


\ifx \showCODEN    \undefined \def \showCODEN     #1{\unskip}     \fi
\ifx \showDOI      \undefined \def \showDOI       #1{#1}\fi
\ifx \showISBNx    \undefined \def \showISBNx     #1{\unskip}     \fi
\ifx \showISBNxiii \undefined \def \showISBNxiii  #1{\unskip}     \fi
\ifx \showISSN     \undefined \def \showISSN      #1{\unskip}     \fi
\ifx \showLCCN     \undefined \def \showLCCN      #1{\unskip}     \fi
\ifx \shownote     \undefined \def \shownote      #1{#1}          \fi
\ifx \showarticletitle \undefined \def \showarticletitle #1{#1}   \fi
\ifx \showURL      \undefined \def \showURL       {\relax}        \fi
\providecommand\bibfield[2]{#2}
\providecommand\bibinfo[2]{#2}
\providecommand\natexlab[1]{#1}
\providecommand\showeprint[2][]{arXiv:#2}

\bibitem[Alkaissi and McFarlane(2023)]%
        {Alkaissi2023ArtificialWriting}
\bibfield{author}{\bibinfo{person}{Hussam Alkaissi} {and} \bibinfo{person}{Samy~I McFarlane}.} \bibinfo{year}{2023}\natexlab{}.
\newblock \showarticletitle{{Artificial Hallucinations in ChatGPT: Implications in Scientific Writing}}.
\newblock \bibinfo{journal}{\emph{Cureus}} (\bibinfo{date}{2} \bibinfo{year}{2023}).
\newblock
\showISSN{2168-8184}
\urldef\tempurl%
\url{https://doi.org/10.7759/cureus.35179}
\showDOI{\tempurl}


\bibitem[Bender et~al\mbox{.}(2021)]%
        {Bender2021OnBig}
\bibfield{author}{\bibinfo{person}{Emily~M Bender}, \bibinfo{person}{Timnit Gebru}, \bibinfo{person}{Angelina McMillan-Major}, {and} \bibinfo{person}{Shmargaret Shmitchell}.} \bibinfo{year}{2021}\natexlab{}.
\newblock \showarticletitle{{On the Dangers of Stochastic Parrots: Can Language Models Be Too Big?}}. In \bibinfo{booktitle}{\emph{Proceedings of the 2021 ACM Conference on Fairness, Accountability, and Transparency}} \emph{(\bibinfo{series}{FAccT '21})}. \bibinfo{publisher}{Association for Computing Machinery}, \bibinfo{address}{New York, NY, USA}, \bibinfo{pages}{610--623}.
\newblock
\showISBNx{9781450383097}
\urldef\tempurl%
\url{https://doi.org/10.1145/3442188.3445922}
\showDOI{\tempurl}


\bibitem[Besan{\c{c}}on et~al\mbox{.}(2021)]%
        {Besancon2021DefinitelyLiterature}
\bibfield{author}{\bibinfo{person}{Lonni Besan{\c{c}}on}, \bibinfo{person}{Yvonne Jansen}, \bibinfo{person}{Andy Cockburn}, {and} \bibinfo{person}{Pierre Dragicevic}.} \bibinfo{year}{2021}\natexlab{}.
\newblock \bibinfo{booktitle}{\emph{{Definitely Maybe: Hedges And Boosters in the HCI Literature; Definitely Maybe: Hedges And Boosters in the HCI Literature}}}.
\newblock \bibinfo{type}{{T}echnical {R}eport}.
\newblock
\urldef\tempurl%
\url{https://hedgesandboosters.github.io}
\showURL{%
\tempurl}


\bibitem[Brown et~al\mbox{.}(2020)]%
        {brown_language_2020}
\bibfield{author}{\bibinfo{person}{Tom Brown}, \bibinfo{person}{Benjamin Mann}, \bibinfo{person}{Nick Ryder}, \bibinfo{person}{Melanie Subbiah}, \bibinfo{person}{Jared~D Kaplan}, \bibinfo{person}{Prafulla Dhariwal}, \bibinfo{person}{Arvind Neelakantan}, \bibinfo{person}{Pranav Shyam}, \bibinfo{person}{Girish Sastry}, \bibinfo{person}{Amanda Askell}, \bibinfo{person}{Sandhini Agarwal}, \bibinfo{person}{Ariel Herbert-Voss}, \bibinfo{person}{Gretchen Krueger}, \bibinfo{person}{Tom Henighan}, \bibinfo{person}{Rewon Child}, \bibinfo{person}{Aditya Ramesh}, \bibinfo{person}{Daniel Ziegler}, \bibinfo{person}{Jeffrey Wu}, \bibinfo{person}{Clemens Winter}, \bibinfo{person}{Chris Hesse}, \bibinfo{person}{Mark Chen}, \bibinfo{person}{Eric Sigler}, \bibinfo{person}{Mateusz Litwin}, \bibinfo{person}{Scott Gray}, \bibinfo{person}{Benjamin Chess}, \bibinfo{person}{Jack Clark}, \bibinfo{person}{Christopher Berner}, \bibinfo{person}{Sam McCandlish}, \bibinfo{person}{Alec Radford}, \bibinfo{person}{Ilya Sutskever}, {and}
  \bibinfo{person}{Dario Amodei}.} \bibinfo{year}{2020}\natexlab{}.
\newblock \showarticletitle{Language {Models} are {Few}-{Shot} {Learners}}. In \bibinfo{booktitle}{\emph{Advances in {Neural} {Information} {Processing} {Systems}}}, \bibfield{editor}{\bibinfo{person}{H.~Larochelle}, \bibinfo{person}{M.~Ranzato}, \bibinfo{person}{R.~Hadsell}, \bibinfo{person}{M.~F. Balcan}, {and} \bibinfo{person}{H.~Lin}} (Eds.), Vol.~\bibinfo{volume}{33}. \bibinfo{publisher}{Curran Associates, Inc.}, \bibinfo{pages}{1877--1901}.
\newblock
\urldef\tempurl%
\url{https://proceedings.neurips.cc/paper_files/paper/2020/file/1457c0d6bfcb4967418bfb8ac142f64a-Paper.pdf}
\showURL{%
\tempurl}


\bibitem[Buruk(2023)]%
        {Buruk2023AcademicTransparency}
\bibfield{author}{\bibinfo{person}{Oğuz~'Oz' Buruk}.} \bibinfo{year}{2023}\natexlab{}.
\newblock \showarticletitle{{Academic Writing with GPT-3.5 (ChatGPT): Reflections on Practices, Efficacy and Transparency}}. In \bibinfo{booktitle}{\emph{Proceedings of the 26th International Academic Mindtrek Conference}} \emph{(\bibinfo{series}{Mindtrek '23})}. \bibinfo{publisher}{Association for Computing Machinery}, \bibinfo{address}{New York, NY, USA}, \bibinfo{pages}{144--153}.
\newblock
\showISBNx{9798400708749}
\urldef\tempurl%
\url{https://doi.org/10.1145/3616961.3616992}
\showDOI{\tempurl}


\bibitem[Chayka(2023)]%
        {Chayka2023IsArtists}
\bibfield{author}{\bibinfo{person}{Kyle Chayka}.} \bibinfo{year}{2023}\natexlab{}.
\newblock \showarticletitle{{Is A.I. Art Stealing From Artists?}}
\newblock \bibinfo{journal}{\emph{The New Yorker}} (\bibinfo{year}{2023}).
\newblock
\urldef\tempurl%
\url{https://www.newyorker.com/culture/infinite-scroll/is-ai-art-stealing-from-artists}
\showURL{%
\tempurl}


\bibitem[Ding et~al\mbox{.}(2023)]%
        {ding_ultrachat_2023}
\bibfield{author}{\bibinfo{person}{Ning Ding}, \bibinfo{person}{Yulin Chen}, \bibinfo{person}{Bokai Xu}, \bibinfo{person}{Shengding Hu}, \bibinfo{person}{Yujia Qin}, \bibinfo{person}{Zhiyuan Liu}, \bibinfo{person}{Maosong Sun}, {and} \bibinfo{person}{Bowen Zhou}.} \bibinfo{year}{2023}\natexlab{}.
\newblock \bibinfo{title}{{UltraChat}: {A} {Large}-scale {Auto}-generated {Multi}-round {Dialogue} {Data}}.
\newblock
\newblock
\urldef\tempurl%
\url{https://github.com/thunlp/ultrachat}
\showURL{%
\tempurl}
\newblock
\shownote{Publication Title: GitHub repository}.


\bibitem[Dourish(2006)]%
        {Dourish2006ImplicationsDesign}
\bibfield{author}{\bibinfo{person}{Paul Dourish}.} \bibinfo{year}{2006}\natexlab{}.
\newblock \showarticletitle{{Implications for Design}}. In \bibinfo{booktitle}{\emph{Conference on Human Factors in Computing Systems (CHI)}}.
\newblock


\bibitem[Eagle et~al\mbox{.}(2023)]%
        {Eagle2023AGaslighting}
\bibfield{author}{\bibinfo{person}{Tessa Eagle}, \bibinfo{person}{Leya~Breanna Baltaxe-Admony}, \bibinfo{person}{Lee Taber}, {and} \bibinfo{person}{Kathryn~E Ringland}.} \bibinfo{year}{2023}\natexlab{}.
\newblock \showarticletitle{{A Playful Twist on the Peer Review Process and Methodological Gaslighting}}. In \bibinfo{booktitle}{\emph{Extended Abstracts of the 2023 CHI Conference on Human Factors in Computing Systems}} \emph{(\bibinfo{series}{CHI EA '23})}. \bibinfo{publisher}{Association for Computing Machinery}, \bibinfo{address}{New York, NY, USA}.
\newblock
\showISBNx{9781450394222}
\urldef\tempurl%
\url{https://doi.org/10.1145/3544549.3582745}
\showDOI{\tempurl}


\bibitem[H{\"{a}}m{\"{a}}l{\"{a}}inen et~al\mbox{.}(2023)]%
        {Hamalainen2023EvaluatingStudy}
\bibfield{author}{\bibinfo{person}{Perttu H{\"{a}}m{\"{a}}l{\"{a}}inen}, \bibinfo{person}{Mikke Tavast}, {and} \bibinfo{person}{Anton Kunnari}.} \bibinfo{year}{2023}\natexlab{}.
\newblock \showarticletitle{{Evaluating Large Language Models in Generating Synthetic HCI Research Data: A Case Study}}. In \bibinfo{booktitle}{\emph{Proceedings of the 2023 CHI Conference on Human Factors in Computing Systems}} \emph{(\bibinfo{series}{CHI '23})}. \bibinfo{publisher}{Association for Computing Machinery}, \bibinfo{address}{New York, NY, USA}.
\newblock
\showISBNx{9781450394215}
\urldef\tempurl%
\url{https://doi.org/10.1145/3544548.3580688}
\showDOI{\tempurl}


\bibitem[Jiang et~al\mbox{.}(2023)]%
        {jiang_mistral_2023}
\bibfield{author}{\bibinfo{person}{Albert~Q. Jiang}, \bibinfo{person}{Alexandre Sablayrolles}, \bibinfo{person}{Arthur Mensch}, \bibinfo{person}{Chris Bamford}, \bibinfo{person}{Devendra~Singh Chaplot}, \bibinfo{person}{Diego de~las Casas}, \bibinfo{person}{Florian Bressand}, \bibinfo{person}{Gianna Lengyel}, \bibinfo{person}{Guillaume Lample}, \bibinfo{person}{Lucile Saulnier}, \bibinfo{person}{Lélio~Renard Lavaud}, \bibinfo{person}{Marie-Anne Lachaux}, \bibinfo{person}{Pierre Stock}, \bibinfo{person}{Teven~Le Scao}, \bibinfo{person}{Thibaut Lavril}, \bibinfo{person}{Thomas Wang}, \bibinfo{person}{Timothée Lacroix}, {and} \bibinfo{person}{William~El Sayed}.} \bibinfo{year}{2023}\natexlab{}.
\newblock \bibinfo{title}{Mistral {7B}}.
\newblock
\newblock
\urldef\tempurl%
\url{2310.06825}
\showURL{%
\tempurl}
\newblock
\shownote{\_eprint: 2310.06825}.


\bibitem[Linxen et~al\mbox{.}(2021)]%
        {Linxen2021HowCHI}
\bibfield{author}{\bibinfo{person}{Sebastian Linxen}, \bibinfo{person}{Christian Sturm}, \bibinfo{person}{Florian Br{\"{u}}hlmann}, \bibinfo{person}{Vincent Cassau}, \bibinfo{person}{Klaus Opwis}, {and} \bibinfo{person}{Katharina Reinecke}.} \bibinfo{year}{2021}\natexlab{}.
\newblock \showarticletitle{{How WEIRD is CHI?}}. In \bibinfo{booktitle}{\emph{Proceedings of the 2021 CHI Conference on Human Factors in Computing Systems}} \emph{(\bibinfo{series}{CHI '21})}. \bibinfo{publisher}{Association for Computing Machinery}, \bibinfo{address}{New York, NY, USA}.
\newblock
\showISBNx{9781450380966}
\urldef\tempurl%
\url{https://doi.org/10.1145/3411764.3445488}
\showDOI{\tempurl}


\bibitem[Marshall and Linehan(2017)]%
        {Marshall2017MisrepresentationLiterature}
\bibfield{author}{\bibinfo{person}{Joe Marshall} {and} \bibinfo{person}{Conor Linehan}.} \bibinfo{year}{2017}\natexlab{}.
\newblock \showarticletitle{{Misrepresentation of Health Research in Exertion Games Literature}}. In \bibinfo{booktitle}{\emph{Proceedings of the 2017 CHI Conference on Human Factors in Computing Systems}} \emph{(\bibinfo{series}{CHI '17})}. \bibinfo{publisher}{Association for Computing Machinery}, \bibinfo{address}{New York, NY, USA}, \bibinfo{pages}{4899--4910}.
\newblock
\showISBNx{9781450346559}
\urldef\tempurl%
\url{https://doi.org/10.1145/3025453.3025691}
\showDOI{\tempurl}


\bibitem[Marshall et~al\mbox{.}(2017)]%
        {Marshall2017ThrowawayResearch}
\bibfield{author}{\bibinfo{person}{Joe Marshall}, \bibinfo{person}{Conor Linehan}, \bibinfo{person}{Jocelyn Spence}, {and} \bibinfo{person}{Stefan Rennick~Egglestone}.} \bibinfo{year}{2017}\natexlab{}.
\newblock \showarticletitle{{Throwaway Citation of Prior Work Creates Risk of Bad HCI Research}}. In \bibinfo{booktitle}{\emph{Proceedings of the 2017 CHI Conference Extended Abstracts on Human Factors in Computing Systems}} \emph{(\bibinfo{series}{CHI EA '17})}. \bibinfo{publisher}{Association for Computing Machinery}, \bibinfo{address}{New York, NY, USA}, \bibinfo{pages}{827--836}.
\newblock
\showISBNx{9781450346566}
\urldef\tempurl%
\url{https://doi.org/10.1145/3027063.3052751}
\showDOI{\tempurl}


\bibitem[Myklebust(2023)]%
        {Myklebust2023UniversitiesAhead}
\bibfield{author}{\bibinfo{person}{Jan~Petter Myklebust}.} \bibinfo{year}{2023}\natexlab{}.
\newblock \bibinfo{title}{{Universities adjust to ChatGPT, but the ‘real AI’ lies ahead}}.
\newblock
\newblock
\urldef\tempurl%
\url{https://www.universityworldnews.com/post.php?story=20230301105802395}
\showURL{%
\tempurl}


\bibitem[Nacke(2019)]%
        {Nacke2019HowEdition}
\bibfield{author}{\bibinfo{person}{Lennart~E. Nacke}.} \bibinfo{year}{2019}\natexlab{}.
\newblock \showarticletitle{{How to write Chi papers (Third edition)}}. In \bibinfo{booktitle}{\emph{Conference on Human Factors in Computing Systems - Proceedings}}. \bibinfo{publisher}{Association for Computing Machinery}.
\newblock
\showISBNx{9781450359719}
\urldef\tempurl%
\url{https://doi.org/10.1145/3290607.3298817}
\showDOI{\tempurl}


\bibitem[Paszke et~al\mbox{.}(2019)]%
        {paszke_pytorch_2019}
\bibfield{author}{\bibinfo{person}{Adam Paszke}, \bibinfo{person}{Sam Gross}, \bibinfo{person}{Francisco Massa}, \bibinfo{person}{Adam Lerer}, \bibinfo{person}{James Bradbury}, \bibinfo{person}{Gregory Chanan}, \bibinfo{person}{Trevor Killeen}, \bibinfo{person}{Zeming Lin}, \bibinfo{person}{Natalia Gimelshein}, \bibinfo{person}{Luca Antiga}, \bibinfo{person}{Alban Desmaison}, \bibinfo{person}{Andreas Köpf}, \bibinfo{person}{Edward Yang}, \bibinfo{person}{Zach DeVito}, \bibinfo{person}{Martin Raison}, \bibinfo{person}{Alykhan Tejani}, \bibinfo{person}{Sasank Chilamkurthy}, \bibinfo{person}{Benoit Steiner}, \bibinfo{person}{Lu Fang}, \bibinfo{person}{Junjie Bai}, {and} \bibinfo{person}{Soumith Chintala}.} \bibinfo{year}{2019}\natexlab{}.
\newblock \showarticletitle{{PyTorch}: {An} {Imperative} {Style}, {High}-{Performance} {Deep} {Learning} {Library}}. In \bibinfo{booktitle}{\emph{Proceedings of the 33rd {International} {Conference} on {Neural} {Information} {Processing} {Systems}}}. \bibinfo{publisher}{Curran Associates Inc.}, \bibinfo{address}{Red Hook, NY, USA}.
\newblock


\bibitem[Pedregosa et~al\mbox{.}(2011)]%
        {pedregosa_scikit-learn_2011}
\bibfield{author}{\bibinfo{person}{F. Pedregosa}, \bibinfo{person}{G. Varoquaux}, \bibinfo{person}{A. Gramfort}, \bibinfo{person}{V. Michel}, \bibinfo{person}{B. Thirion}, \bibinfo{person}{O. Grisel}, \bibinfo{person}{M. Blondel}, \bibinfo{person}{P. Prettenhofer}, \bibinfo{person}{R. Weiss}, \bibinfo{person}{V. Dubourg}, \bibinfo{person}{J. Vanderplas}, \bibinfo{person}{A. Passos}, \bibinfo{person}{D. Cournapeau}, \bibinfo{person}{M. Brucher}, \bibinfo{person}{M. Perrot}, {and} \bibinfo{person}{E. Duchesnay}.} \bibinfo{year}{2011}\natexlab{}.
\newblock \showarticletitle{Scikit-learn: {Machine} {Learning} in {Python}}.
\newblock \bibinfo{journal}{\emph{Journal of Machine Learning Research}}  \bibinfo{volume}{12} (\bibinfo{year}{2011}), \bibinfo{pages}{2825--2830}.
\newblock


\bibitem[Penny~Light et~al\mbox{.}(2015)]%
        {penny_light_feminist_2015}
\bibfield{author}{\bibinfo{person}{Tracy Penny~Light}, \bibinfo{person}{Jane Nicholas}, {and} \bibinfo{person}{Renee Bondy}.} \bibinfo{year}{2015}\natexlab{}.
\newblock \bibinfo{booktitle}{\emph{Feminist {Pedagogy} in {Higher} {Education}: {Critical} {Theory} and {Practice}}}.
\newblock
\showISBNx{978-1-77112-114-9}


\bibitem[Pohl and Mottelson(2019)]%
        {Pohl2019HowCHI}
\bibfield{author}{\bibinfo{person}{Henning Pohl} {and} \bibinfo{person}{Aske Mottelson}.} \bibinfo{year}{2019}\natexlab{}.
\newblock \showarticletitle{{How we guide, write, and cite at CHI}}. In \bibinfo{booktitle}{\emph{Conference on Human Factors in Computing Systems - Proceedings}}. \bibinfo{publisher}{Association for Computing Machinery}.
\newblock
\showISBNx{9781450359719}
\urldef\tempurl%
\url{https://doi.org/10.1145/3290607.3310429}
\showDOI{\tempurl}


\bibitem[Radford et~al\mbox{.}(2018)]%
        {radford_improving_2018}
\bibfield{author}{\bibinfo{person}{Alec Radford}, \bibinfo{person}{Karthik Narasimhan}, \bibinfo{person}{Tim Salimans}, {and} \bibinfo{person}{Ilya Sutskever}.} \bibinfo{year}{2018}\natexlab{}.
\newblock \bibinfo{booktitle}{\emph{Improving {Language} {Understanding} by {Generative} {Pre}-{Training}}}.
\newblock \bibinfo{type}{{T}echnical {R}eport}. \bibinfo{institution}{OpenAI}.
\newblock


\bibitem[Ramesh et~al\mbox{.}(2022)]%
        {ramesh_hierarchical_2022}
\bibfield{author}{\bibinfo{person}{Aditya Ramesh}, \bibinfo{person}{Prafulla Dhariwal}, \bibinfo{person}{Alex Nichol}, \bibinfo{person}{Casey Chu}, {and} \bibinfo{person}{Mark Chen}.} \bibinfo{year}{2022}\natexlab{}.
\newblock \bibinfo{title}{Hierarchical {Text}-{Conditional} {Image} {Generation} with {CLIP} {Latents}}.
\newblock
\newblock
\newblock
\shownote{\_eprint: 2204.06125}.


\bibitem[Rienecker et~al\mbox{.}(2015)]%
        {2015UniversityLearning}
\bibfield{editor}{\bibinfo{person}{Lotte Rienecker}, \bibinfo{person}{Peter~Stray J{\o}rgensen}, \bibinfo{person}{Jens Dolin}, {and} \bibinfo{person}{Gitte~Holten Ingerslev}} (Eds.). \bibinfo{year}{2015}\natexlab{}.
\newblock \bibinfo{booktitle}{\emph{{University Teaching and Learning}}}.
\newblock \bibinfo{publisher}{Samfundslitteratur}. 0--496 pages.
\newblock
\showISBNx{8759319607, 9788759319604}


\bibitem[Touvron et~al\mbox{.}(2023)]%
        {touvron_llama_2023}
\bibfield{author}{\bibinfo{person}{Hugo Touvron}, \bibinfo{person}{Thibaut Lavril}, \bibinfo{person}{Gautier Izacard}, \bibinfo{person}{Xavier Martinet}, \bibinfo{person}{Marie-Anne Lachaux}, \bibinfo{person}{Timothée Lacroix}, \bibinfo{person}{Baptiste Rozière}, \bibinfo{person}{Naman Goyal}, \bibinfo{person}{Eric Hambro}, \bibinfo{person}{Faisal Azhar}, \bibinfo{person}{Aurelien Rodriguez}, \bibinfo{person}{Armand Joulin}, \bibinfo{person}{Edouard Grave}, {and} \bibinfo{person}{Guillaume Lample}.} \bibinfo{year}{2023}\natexlab{}.
\newblock \bibinfo{title}{{LLaMA}: {Open} and {Efficient} {Foundation} {Language} {Models}}.
\newblock
\newblock
\urldef\tempurl%
\url{2302.13971}
\showURL{%
\tempurl}
\newblock
\shownote{\_eprint: 2302.13971}.


\bibitem[Tunstall et~al\mbox{.}(2023)]%
        {tunstall_zephyr_2023}
\bibfield{author}{\bibinfo{person}{Lewis Tunstall}, \bibinfo{person}{Edward Beeching}, \bibinfo{person}{Nathan Lambert}, \bibinfo{person}{Nazneen Rajani}, \bibinfo{person}{Kashif Rasul}, \bibinfo{person}{Younes Belkada}, \bibinfo{person}{Shengyi Huang}, \bibinfo{person}{Leandro~von Werra}, \bibinfo{person}{Clémentine Fourrier}, \bibinfo{person}{Nathan Habib}, \bibinfo{person}{Nathan Sarrazin}, \bibinfo{person}{Omar Sanseviero}, \bibinfo{person}{Alexander~M. Rush}, {and} \bibinfo{person}{Thomas Wolf}.} \bibinfo{year}{2023}\natexlab{}.
\newblock \bibinfo{title}{Zephyr: {Direct} {Distillation} of {LM} {Alignment}}.
\newblock
\newblock
\urldef\tempurl%
\url{2310.16944}
\showURL{%
\tempurl}
\newblock
\shownote{\_eprint: 2310.16944}.


\bibitem[Tyack and Mekler(2020)]%
        {Tyack2020Self-DeterminationQuestions}
\bibfield{author}{\bibinfo{person}{April Tyack} {and} \bibinfo{person}{Elisa~D. Mekler}.} \bibinfo{year}{2020}\natexlab{}.
\newblock \showarticletitle{{Self-Determination Theory in HCI Games Research: Current Uses and Open Questions}}.
\newblock  (\bibinfo{year}{2020}), \bibinfo{pages}{1--22}.
\newblock
\showISBNx{9781450367080}
\urldef\tempurl%
\url{https://doi.org/10.1145/3313831.3376723}
\showDOI{\tempurl}


\end{thebibliography}


\end{document}